\documentclass[conference]{IEEEtran}
\pdfoutput=1

\usepackage[pdflang={en-US},pdftex]{hyperref}
\usepackage[]{bookmark}
\usepackage{balance}       
\usepackage{graphics}      
\usepackage[T1]{fontenc}   
\usepackage{txfonts}
\usepackage{mathptmx}

\usepackage{color}
\usepackage{booktabs}
\usepackage{textcomp}

\usepackage{microtype}        
\usepackage{ccicons}          

\usepackage{todonotes}
\usepackage{float}

\def\plaintitle{Handwritten Signature Verification Using Hand-Worn Devices}

\def\emptyauthor{}
\def\plainkeywords{Machine Learning; Internet of Wearable Things; Intoxication Detection; Smart Watch; Google Glass; Sensors }

\makeatletter
\def\url@leostyle{%
  \@ifundefined{selectfont}{
    \def\UrlFont{\sf}
  }{
    \def\UrlFont{\small\bf\ttfamily}
  }}
\makeatother
\def\pprw{8.5in}
\def\pprh{11in}

\definecolor{linkColor}{RGB}{6,125,233}
\hypersetup{%
  pdftitle={\plaintitle},
  pdfauthor={\emptyauthor},
  pdfkeywords={\plainkeywords},
  pdfdisplaydoctitle=true, 
  bookmarksnumbered,
  pdfstartview={FitH},
  colorlinks,
  citecolor=black,
  filecolor=black,
  linkcolor=black,
  urlcolor=linkColor,
  breaklinks=true,
  hypertexnames=false
}

\begin{document}

\title{\plaintitle}

\author{\IEEEauthorblockN{Ben Nassi\IEEEauthorrefmark{1},
Alona Levy\IEEEauthorrefmark{2}, Yuval Elovici\IEEEauthorrefmark{1} Erez Shmueli\IEEEauthorrefmark{2}
}
\IEEEauthorblockA{\IEEEauthorrefmark{1}	Dept. of Software and Information Systems Engineering, Ben-Gurion University of the Negev, Israel\\ 
\IEEEauthorrefmark{2}	Dept. of Industrial Engineering, Tel-Aviv University , Israel\\ 
Email: nassib@post.bgu.ac.il, alonale1@mail.tau.ac.il, elovici@bgu.ac.il, shmueli@tau.ac.il
}}
\maketitle

\begin{abstract}
Online signature verification technologies, such as those available in banks and post offices, rely on dedicated digital devices such as tablets or smart pens to capture, analyze and verify signatures.
In this paper, we suggest a novel method for online signature verification that relies on the increasingly available hand-worn devices, such as smartwatches or fitness trackers, instead of dedicated ad-hoc devices.
Our method uses a set of known genuine and forged signatures, recorded using the motion sensors of a hand-worn device, to train a machine learning classifier.
Then, given the recording of an unknown signature and a claimed identity, the classifier can determine whether the signature is genuine or forged.
In order to validate our method, it was applied on 1980 recordings of genuine and forged signatures that we collected from 66 subjects in our institution.
Using our method, we were able to successfully distinguish between genuine and forged signatures with a high degree of accuracy (0.98 AUC and 0.05 EER). 
\end{abstract}

\section{Introduction}

Financial fraud is a common occurrence across the globe, causing a significant amount of damage to the economy.
According to recent surveys of consumer fraud \cite{fraud}, the Federal Trade Commission estimated that 37.8 million incidents of fraud took place in 2011 in the US only and the Financial Fraud Research Center estimated that \$40 to \$50 billion is lost to fraud annually.

Despite the prevention efforts of banks, businesses and the law enforcement community, 
according to a 2012 survey by the Association for Financial Professionals \cite{afp}, paper checks continue to lead as the payment type most susceptible to fraudulent attacks and as the payment method accounting for the largest dollar amount of loss due to fraud.
According to another survey \cite{aba}, in 2011 alone, the American Bankers Association estimated the use of paper checks in 34 trillion dollars, and the losses due to check fraud in 1.2 billion dollars. 

Paper checks as well as other legal, financial and administrative documents rely on the handwritten signature as an important behavioral trait to verify a person's identify.
One of the main reasons for its widespread use is that the process of collecting handwritten signatures is non-invasive and familiar, given that people routinely use signatures in their daily life \cite{plamondon2000online}.

In a typical handwritten signature verification system, a user claims to be a particular individual, and provides a sample of her signature.
The role of the verification system is to determine, based on the signature sample, whether the user is indeed who he/she claims to be.

Depending on the data acquisition type, signature verification methods can be classified into two approaches: the \emph{offline} approach relies on the static handwriting image and the \emph{online} approach relies on the dynamic trajectory of the pen tip.
While the latter approach usually requires a designated ad hoc device (commonly called a digitizer), the additional time dimension provides valuable information about the signature, therefore leading to a higher verification performance in general \cite{impedovo2008automatic}.

In this paper, we suggest a new approach for signature verification that is based on data acquired from hand-worn devices.
Hand-worn devices, such as smartwatches and fitness trackers, are becoming increasingly adopted by consumers, and according to recent reports, one out of every six people in the US already use a smartwatch \cite{wearables2016} and the overall smartwatch market is expected to reach 373 million devices by 2020.
We hypothesize that it is possible to verify handwritten signatures accurately by analyzing motion data (i.e., accelerometer and gyroscope measurements) collected from hand-worn devices.
We base our hypothesis on the assumption that people adopt a specific signing pattern over the years that is (1) unique and very difficult for others to imitate, and (2) this uniqueness can be captured adequately using the motion sensors of a hand-worn device.

Our approach attempts to combine the benefits of both the offline and online verification approaches.
Similar to the offline approach, our approach does not require a designated ad hoc device to capture the signature.
The collection of the signature itself can take place on a regular sheet of paper (as is the case of many types of contracts, receipts and other documents that have not yet been digitized) using a hand-worn device.
Like the online verification approach, our approach is able to comprehensively capture the dynamics of the signing process.

Following this approach, we develop a concrete method for online signature verification based on motion data collected from hand-worn devices.
We address the signature verification task as a machine learning classification problem.
Our method uses a set of known genuine and forged signatures, recorded using the motion sensors of a hand-worn device, to train a machine learning classifier.
This model learns the indicators that allow to distinguish between genuine and forged signatures.
It is important to note that we use a single global model that was learned from a training dataset comprising a relatively small set of users.
In other words, we do not generate a personalized classification model for each user.
Then, given the recording of an unknown signature and a claimed identity, our classification model can determine whether the signature is genuine or forged.

In order to validate our method, it was applied on 1980 recordings of genuine and forged signatures that we collected from 66 subjects in our institution.
Using our method, we were able to successfully distinguish between genuine and forged signatures with a high degree of accuracy (0.98 AUC and 0.05 EER). 

The rest of this paper is structured as follows:
In section \ref{sec:relatedwork} we provide the relevant background and list related work.
Section \ref{sec:method} outlines the proposed method.
In section \ref{sec:dataset} we describe the experiment we conducted to collect genuine and forged signatures.
In section \ref{sec:evaluation} we detail the evaluation of our system.
In section \ref{sec:deployment} we discuss a potential deployment of our system.
Section \ref{sec:summary} summarizes the paper and proposes directions for future work.

\section{Background \& Related Work}
\label{sec:relatedwork}
In this section, we provide the relevant background, describing the main concepts and methods of two related areas of study: (1) handwritten signature verification and (2) authentication via wearable devices.

\subsection{Handwritten Signature Verification}
Signature verification systems aim to automatically classify query signatures as genuine (i.e. confirm that they were signed by the claimed user) or forged.
Such systems usually consist of an enrollment phase, during which a system's user provides samples of his/her signature, and an operation (or classification) phase, in which the user claims the identity of a person and provides a query signature.
The system then classifies such a query signature as either genuine or a forgery.

Depending on the data acquisition type, signature verification methods can be classified as online (dynamic) or offline (static).
Traditional signature verification methods are based on the offline handwriting image.
In this case, the signature is represented as a digital image, usually in grayscale format, comprising of a set of points $S(x, y)$; $0 \leq x \leq H$; $0 \leq y \leq W$, where $H$ and $W$ denote the height and width of the image.

In contrast, online signature verification methods take the dynamic writing process into account \cite{plamondon2000online}.
A signature is represented by a pen tip trajectory measurement that captures the position of the pen over time; depending on the digitizer, this may be accompanied by additional measurements of the pressure and pen inclination.
In this case, the signature is represented as a sequence $S(n)$; $n = 1,\ldots,N$, where $S(n)$ is the signal sampled at time $n \cdot \Delta t$ and $\Delta t$ is the sampling interval \cite{hafemann2015offline}.
Clearly, the additional time dimension captured by online methods provides valuable information about the signature, leading to a higher level of verification performance in general \cite{impedovo2008automatic}.

Our approach attempts to combine the benefits of both the offline and online verification approaches.
Similar to the offline approach, our approach does not require a designated ad-hoc device to capture the signature.
The collection of the signature itself can take place on a regular sheet of paper (as is the case of many types of contracts, receipts and other documents that have not yet been digitized) using a hand-worn device.
Like the online verification approach, our approach is able to comprehensively capture the dynamics of the signing process.

Two approaches to online signature verification can be further distinguished.
Feature-based methods represent signatures with feature vectors while function-based methods take the complete time sequence into account \cite{plamondon1989automatic}.
The former provide a data security advantage because the original signatures do not have to be stored in a database; however,  the latter tend to achieve better verification performance.

State-of-the-art methods for function-based verification include hidden Markov models (HMM) \cite{fierrez2007hmm} and dynamic time warping (DTW) \cite{kholmatov2005identity}.
HMMs are statistical models that require a considerable number of reference signatures per user for training. 
In contrast, DTW matches signatures directly with reference samples of the claimed user and is particularly useful if only a few reference signatures are available, which is a typical scenario.
More specifically, DTW computes a dissimilarity score between two time sequences.
Taking into account the (possibly different) lengths of the two sequences, the sequences are aligned along a common time axis such that the sum of Euclidean distances between the feature vectors along the warping path is minimal.
With regard to signatures, DTW matches two signatures by aligning the pen-tip trajectory measurements along a common time axis.
The resulting distance depends on the sequence length of the two signatures and needs to be compared with a threshold in order to accept or reject the claimed identity.

Several variations of the function-based methods use a Discrete Cosine Transform (DCT) or a Discrete Fourier Transform (DFT) compression of the signal instead of using its raw form.
While mainly used in the field of speech recognition \cite{nirjon2013smfcc, rao2014discrete}, some papers have tested the effect of using DCT and DFT in signature verification systems.
In \cite{rashidi2012feature} the authors used DCT coefficients as features with which to train an online signature verification model, without combining it with DTW, and obtained satisfactory results.
In another paper \cite{yanikoglu2009online} the combined use of Fourier descriptors along with DTW was shown to improve the verification results when compared to those obtained by simply using DTW on its own. 

In this paper we propose a verification method that combines the function-based approach (using DCT and DTW) and the feature-based approach.

\begin{figure*}
\centering
\includegraphics[width=1.0\textwidth]{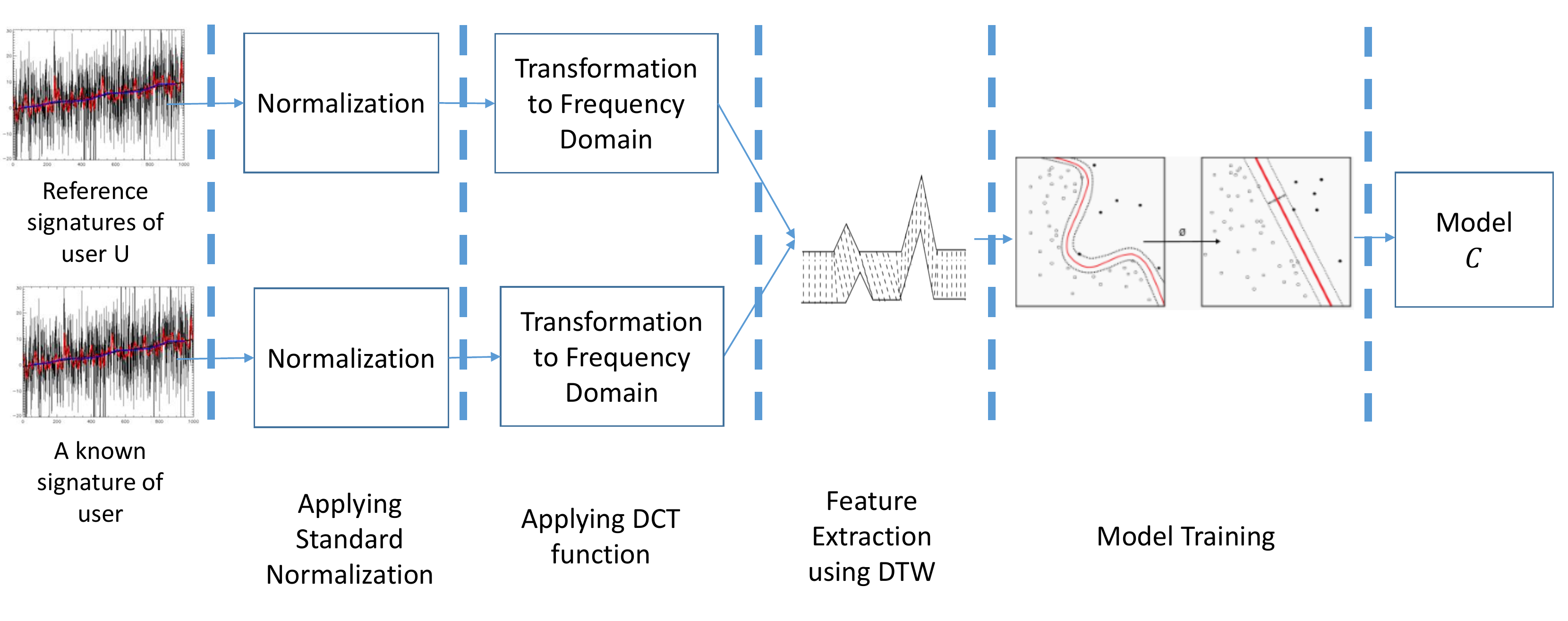}
\caption{The training phase.}
\label{fig:steps}
\end{figure*}

\subsection{User Authentication via Wearable Devices}

A variety of recent works suggested the use of wearable devices for the tasks of user authentication and gesture recognition.
Most of these works rely on the motion sensors (typically accelerometer and gyroscope) embedded in these devices to detect and understand unique movements of the wearing person.

Wrist-worn devices, such as smartwatches and fitness trackers, have become perhaps the most popular category of wearable devices, and many major manufacturers, including Samsung and Apple, have released their devices recently.
Since these devices are worn on the wrist, they introduce a unique opportunity to both detect and understand a user's arm, hand and finger movements as shown in \cite{xu2015finger}.
In comparison, forearm devices such as the one presented in \cite{myo2016}, are very limited in detecting fine gestures such as finger gestures or writing.
Similarly, finger worn devices such as the one introduced in \cite{gummeson2014energy}, can be used to understand users' finger gestures and writing.
However, this is limited to the gestures of a specific finger, and gestures using other fingers can not be identified.
Wrist-worn devices are less limited as they facilitate gesture recognition based on the arm, the hand and all of the fingers.

While there has been a lot of research in the field of user authentication using smartphone devices, there have been only a few works that aimed to authenticate users using wearable devices.
For example, a recent study showed that it is possible to distinguish between users who use the same objects (e.g., a light switch, a refrigerator, etc.) \cite{ranjan2015object} using continuous authentication.
Another research \cite{kumar2016authenticating} aimed at authenticating users from a short recording of their natural walk as captured by their smartwatch.
A recent patent \cite{shin2016wearable} filed by Samsung suggests a novel method by which the veins of a smartwatch user are used to authenticate his/her identity.

Closer to the field of handwriting analysis, several recent studies have tried to use motion data collected from wearable devices to recognize different writing gestures such as inferring the letter written.
For example, the authors of \cite{agrawal2011using} investigated the task of writing in the air with a mobile device. 
In \cite{arduserrecognizing}, researchers suggested a platform for recognizing text written on a whiteboard using a smartwatch.
In \cite{xu2015finger}, researchers tried to infer letters written on a sheet of paper.
With a totally different purpose in mind, the authors of \cite{wang2015mole}, tried to detect the letters typed on a keyboard using a smartwatch.
Similarly, in \cite{beltramelli2015deep}, researchers presented a new attack method that allows attackers to extract sensitive information such as credit card or phone access PIN codes from motion sensors in wearable devices.

However, to the best of our knowledge, none of the existing studies have addressed the task of handwritten signature verification using motion data collected from wearable devices in general and wrist-worn devices in particular.

\section{The Proposed Method}
\label{sec:method}

We hypothesize that it is possible to verify handwritten signatures accurately by analyzing motion data (i.e., accelerometer and gyroscope measurements) collected from hand-worn devices.
We base our hypothesis on the assumption that most people adopt a specific signing pattern over the years that is (1) unique and very difficult for others to imitate, and (2) this uniqueness can be captured adequately using the motion sensors of a hand-worn device.

We address the signature verification task as a machine learning classification problem.
Our method uses a set of known genuine and forged signatures, recorded using the motion sensors of a hand-worn device, to train a machine learning classifier.
This model learns the indicators that allow to distinguish between genuine and forged signatures.
Then, given the recording of an unknown signature and a claimed identity, the classifier can determine whether the signature is genuine or forged.

It is important to note that we use a single global model that was learned from a training dataset comprising a relatively small set of users.
In other words, we do not generate a personalized classification model for each user.
This global model is than applied on new users (that were not part of the training dataset) to classify signatures into genuine or forged.

We now describe the training phase and the enrollment and operation phases of the verification process in details.

\begin{figure*}
\centering
\includegraphics[width=1.0\textwidth]{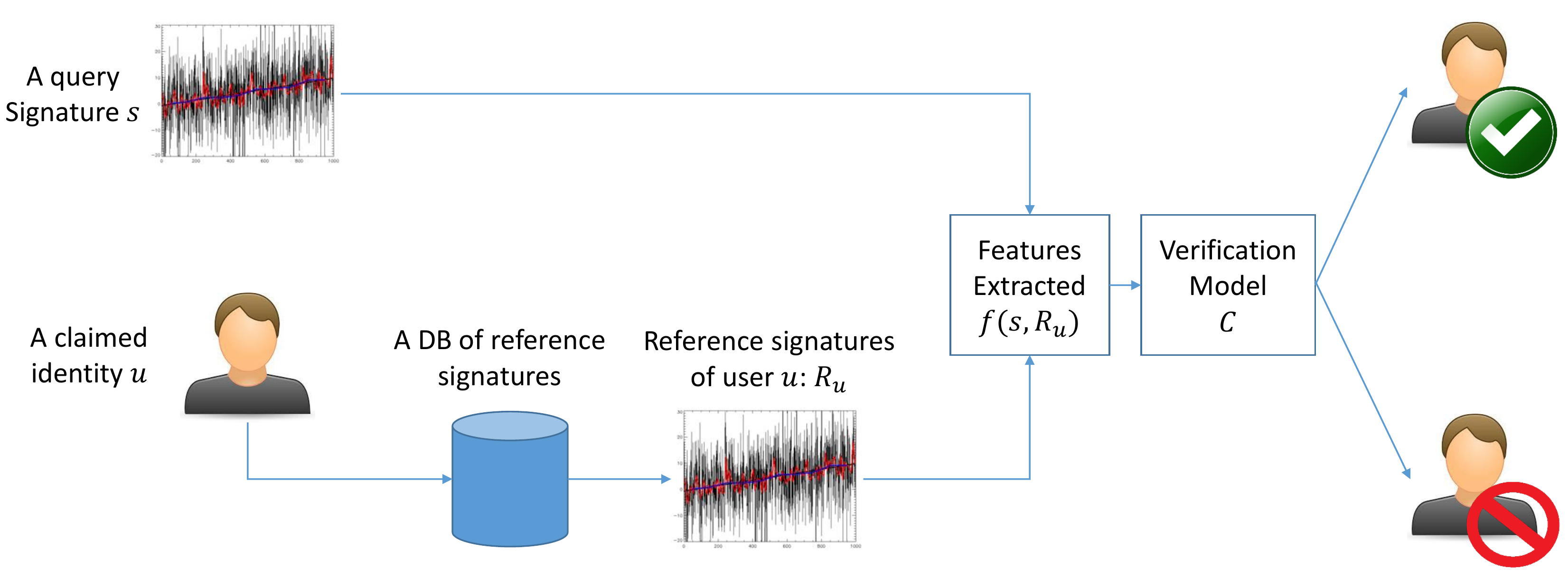}
\caption{The verification phase.}
\label{fig:verification}
\end{figure*}

\subsection{The Training Phase}

During the training phase, we assume that for each enrolled user $u$ we have a known set of genuine signatures $G_u$ and a known set of forged signatures $F_u$.

From the set of genuine signatures $G_u$ of user $u$, we randomly select a subset of genuine signatures to serve as her reference signatures, denoted by $R_u$.
The remaining genuine signatures in $G_u-R_u$ are denoted as $\overline{G_u}$.

For each user $u$, for each signature $s$ in $\overline{G_u} \cup F_u$, we apply the following steps (also illustrated in Figure \ref{fig:steps}):

\begin{enumerate}
\item Normalization - We normalize each motion signal of the signature.
\item Domain transformation - We apply DCT on each of the normalized motion signals to represent them in the frequency domain.
\item Feature Extraction - for each of the normalized motion signals (represented in the transformed domain) we go over each of the reference signature $s_r$ in $R_u$, and calculate the DTW dissimilarity score for the two corresponding motion signals of $s$ and $s_r$. Then we choose the minimal DTW score over all reference signatures.
\item Labeling - We label the instance as GENUINE if $s \in \overline{G_u}$ or as FORGED if $s \in F_u$.
\item Learning a Model - We use all labeled instances to train a machine learning classifier. The trained model is later used in the verification phase to distinguish between unknown genuine and forged signatures.
\end{enumerate}

We now describe in details the normalization, domain transformation and feature extraction stages mentioned above.

\subsubsection{Normalization}

Each of the signatures go through a normalization process in order to enable comparison by means of an Euclidean distance.
Recall that each signature is composed of several (say $D$) motion signals, each of which is a time sequence of real values.
Each motion signal $m_d$, $d=1 \ldots D$, is normalized separately using:

$$\hat{m_d} = \frac{m_d- \mu_{m_d}}{\sigma_{m_d}}$$

where $\mu_{m_d}$ and $\sigma_{m_d}$ are the mean and standard deviation of motion signal $m_d$ respectively.

\subsubsection{Domain Transformation}

We transform each of the normalized motion signals $\hat{m_d}$ from the time domain to the frequency domain by using Discrete Cosine Transform to obtain a compressed representation of the signal, denoted by $c_d$.
We do so by taking the first 20 DCT coefficients to represent each motion signal (recall that the first coefficients retain the most energy of the signal).

\subsubsection{Feature Extraction}
Given a questioned signature $s_i$, and a set of reference signatures $R_u$, for each reference signature $s_r \in R_u$ and for each one of the motion dimensions $d=1,\ldots,D$ we define $dis(s_i, s_r, d)$ as:

$$ dis(s_i, s_r, d) = DTW(c_d^i, c_d^r) $$

where $DTW$ is the dynamic time warping dissimilarity function, $c_d^i$ is the $d$-th normalized and compressed motion signal of the questioned signature $s_i$ and $c_d^r$ is the $d$-th normalized and compressed motion signal of the reference signature $s_r$ .

Finally our vector of extracted features for a signature $s_i$ and the set of reference signatures $R_u$ for the claimed user $u$, is calculated as the minimal $DTW$ distance between $s_i$ and each one of the reference signatures $s_r \in R_u$ (calculated separately for each one of the $D$ motion signal dimensions):

$$ f(s_i, R_u) = <\min_{s_r \in R_u} dis(s_i, s_r, 1), \ldots,  \min_{s_r \in R_u} dis(s_i, s_r, D)>$$

\subsection{The Enrollment and Operation Phases}

Every new (unknown) user $u$ that would like to use the proposed system, has to enroll first by providing its identity and a set of (genuine) reference signatures $R_u$.
This phase is performed only once per user.

Then, given a new (unknown) signature $s$, a claimed identity of an enrolled user $u$ and a trained classifier $C$, we perform the following steps.
First the set of reference signatures $R_u$ for user $u$ is retrieved.
Then, the feature vector $f(s, R_u)$ is calculated as described in the previous subsection.
Finally we apply the classifier $C$ on the tuple $f(s, R_u)$ and return the classification result (GENUINE or FORGED signature).
The operation phase is illustrated in Figure \ref{fig:verification}.

\section{Data Collection}
\label{sec:dataset}

In order to evaluate our method, we conducted an experiment to collect genuine and forged signatures.
In the following subsections we describe the data collection system and the data collection experiment.

\subsection{The Data Collection System}

We used the Microsoft Band \cite{Microsoft} version 1 announced by Microsoft in 2014, and a Samsung GT-N5110 tablet for the experiment.

\begin{figure}[H]
\centering
\includegraphics[width=0.9\columnwidth]{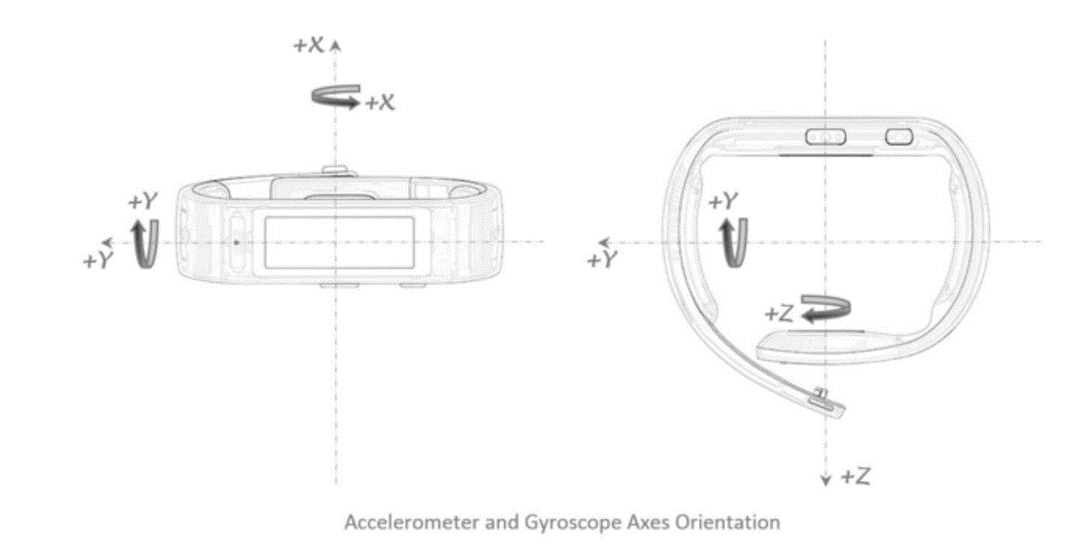}
\caption{The Microsoft Band's sensors and axes. The figure was taken from Microsoft Band SDK Documentation \cite{msbandsdk}}
\label{fig:sensors}
\end{figure}

The Microsoft Band includes both an accelerometer and a gyroscope sensors, and each sample of these sensors includes three different types of measurements:
(1) the acceleration as measured by the accelerometer,
(2) the angle acceleration measured by the gyroscope, and
(3) the angle velocity also measured by the gyroscope.
Each such type of information is provided over three axes ($X$, $Y$ and $Z$), resulting in a total of 9 different dimensions for each sample.
Figure \ref{fig:sensors} depicts the two sensors and the three axes of the Microsoft Band.
These 9 dimensions provide an adequate infrastructure to test our hypothesis about the uniqueness of the hand movement during the signing process. 

The Microsoft Band provides an SDK to communicate with its sensors using an application installed on a paired Bluetooth device.
Therefore, we developed an Android application that was installed on the paired Tablet (Samsung GT-N5110) in order to obtain and record the measurements of its motion sensors during the signing process.
Each sensor was sampled at the maximum rate it supports (62 samples per second).

\subsection{The Data Collection Experiment}
Data collection took place in two phases.
In the first phase, the participants (a class of 66 undergraduate students at our institution) were asked to provide several samples of their genuine signature on a tablet device, using the device's digital pen and while wearing a hand-worn device.
In the second phase, each participant was shown trace recordings of several genuine signatures from the first phase, and was then requested to forge these signatures.
Again, this was done on a tablet device, using the device's digital pen and while wearing the hand-worn device.

All 66 students attended the two sessions, which took place approximately one week apart and together lasted a total of 3 weeks.
Out of the 66 students both sexes were represented fairly equally with a slight female majority of 39 (59\%) female students.
Regarding the participants' dominant hand, out of the 66 there were 57 (86.4\%) right-handed students and only 9 left-handed students (13.6\%).

As an incentive to participate in the experiment, students in the course received 1.5 bonus points to their final course grade for completing the two phases of the experiment.
Moreover, in order to incentivize the participants to provide high quality forgeries, half of the participants - the ones who were ranked among the top 50\% 'best forgerers' - received an additional 0.5 points to their final grade.

The experiment was approved by the ethics committee of our institute.

\subsubsection{Phase 1: genuine signatures}

\begin{figure}[H]
\centering
\includegraphics[width=0.9\columnwidth]{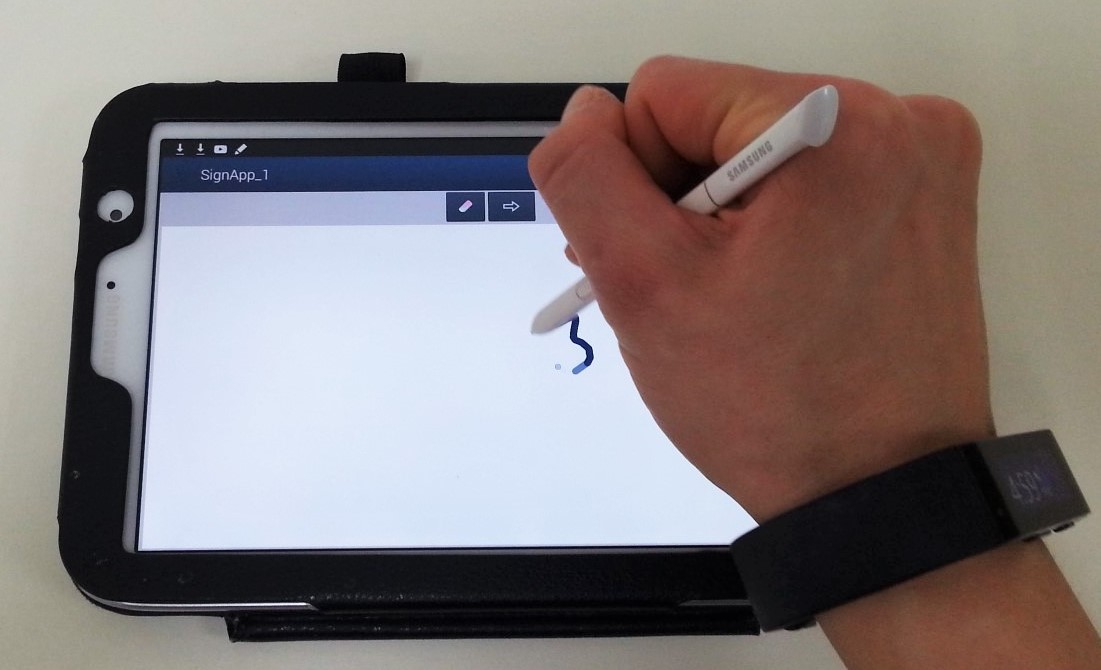}
\caption{The signing process. Each participant was asked to provide 15 samples of their genuine signature.}
\label{fig:signing}
\end{figure}

In this phase, we collected genuine signatures of 66 participants.
Each one of the 66 participants was asked to provide 15 samples of their genuine signature.
They did so by signing on a tablet device, using the device's digital pen and while wearing a hand-worn device.
This process is illustrated in Figure \ref{fig:signing}.

\subsubsection{Phase 2: forged signatures}

\begin{figure}[H]
\centering
\includegraphics[width=0.9\columnwidth]{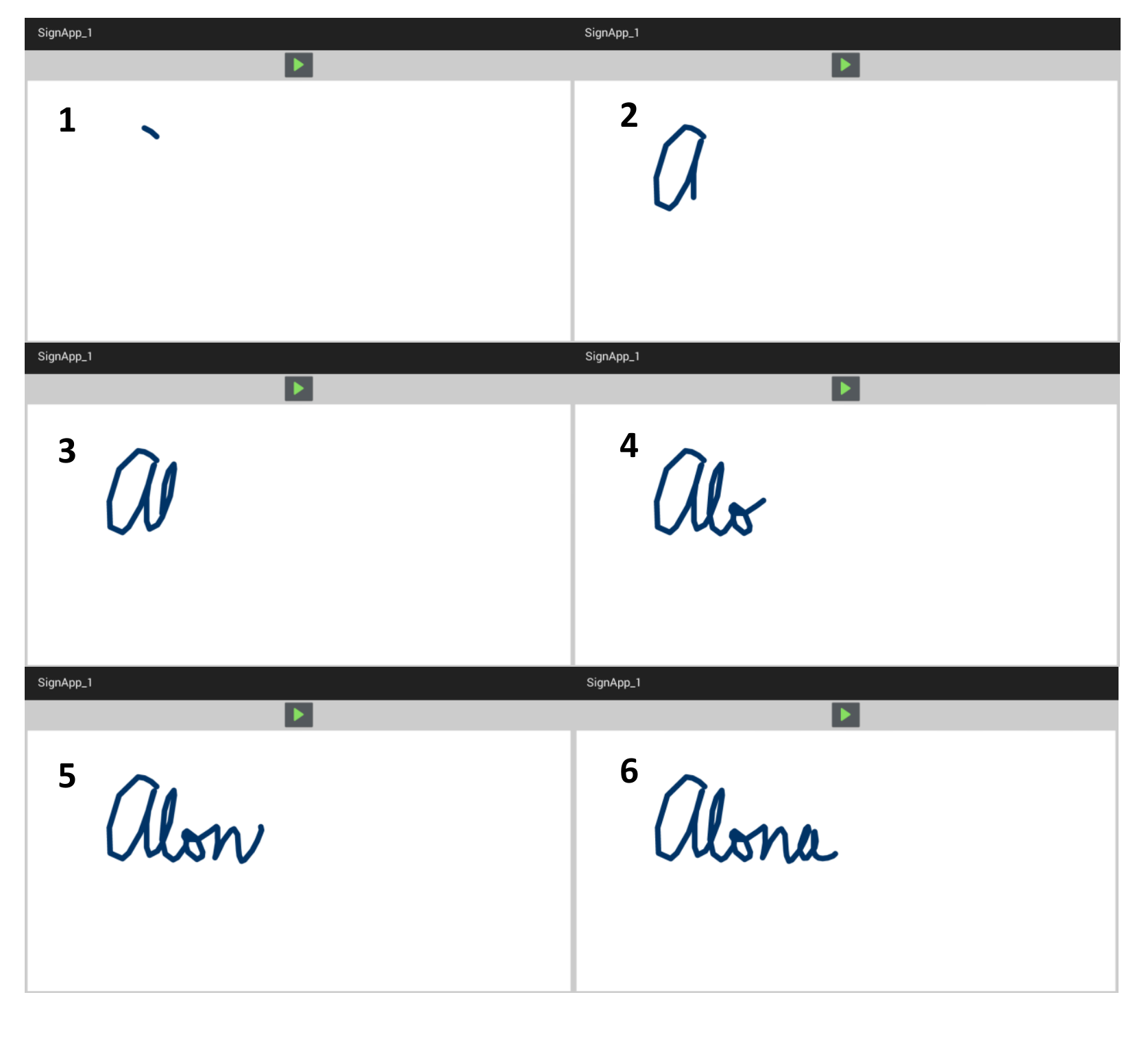}
\caption{Practicing a forgery. For each one of the five signatures the participant had to forge, a video of the signing process was played on the tablet screen in a continuous loop. The figure presents 6 frames taken at various points in time of one such signature.}
\label{fig:forging}
\end{figure}

In the second phase, each participant was asked to forge the signature of five other chosen participants (see details below on how we chose the five participants), that provided samples of their genuine signatures in the first phase.
For each one of these 5 chosen participants, we randomly chose one of his/her 15 genuine signature samples, and simulated its signing trace on the tablet screen (see Figure \ref{fig:forging}).
Each simulation ran in a continuous loop for a time frame of two minutes, and the participants were able to pause/play the simulation at will within this time frame.
Furthermore, participants were instructed to practice tracing the signature as accurately as possible in terms of speed, trajectory etc. during the simulation.

Immediately following each simulation the participants were requested to provide three of their best forgery attempts of the signature they had just seen - each attempt on a fresh blank tablet screen - all while wearing the hand-worn device.
It is worth noting that participants were allowed to erase forgery attempts until they were satisfied that their attempt was good enough to count as one of the 3 submitted attempts.
This ensures us that the skilled forgeries are indeed of high quality and are not affected by the transition from a simulated environment to a non-simulated environment.

In order to ensure that in the second phase, we result in the exact same number of forgeries for each genuine signature, we performed the following procedure.
We first fixed an ordering of the 66 participants (the ordering was randomly generated).
Then, each participant was requested to forge the signatures of the five participants that followed him/her in that fixed ordering (in a cyclic manner).

In summary, our final signatures dataset contains 30 signatures for each one of the 66 participants: 15 genuine samples and 15 forgeries (5 different forgers that provided 3 forgeries each).

\begin{figure}[H]
\centering
\includegraphics[width=0.9\columnwidth]{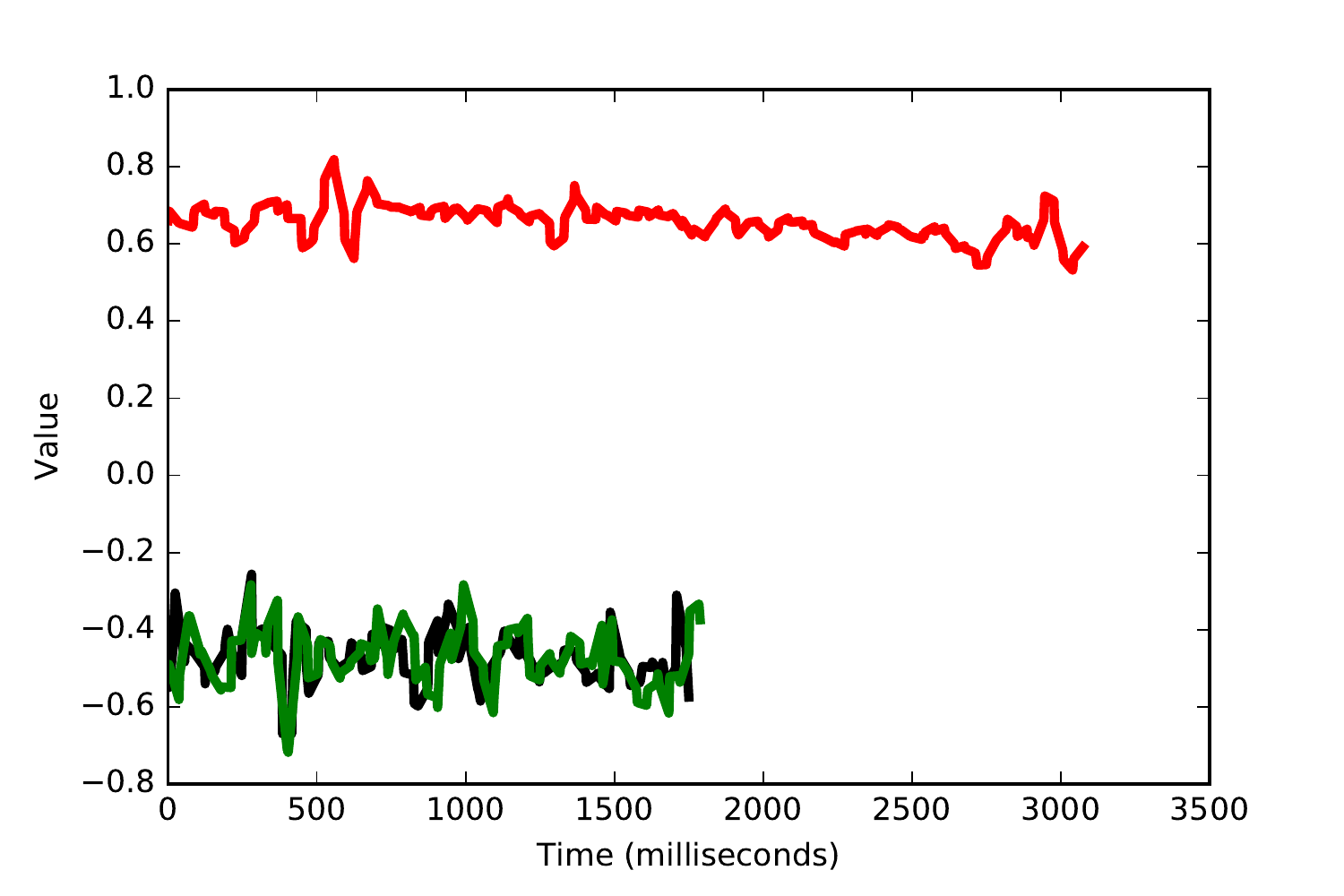}
\caption{An illustration of one signal (accelerometer-$Y$) for a reference signature (black) a genuine signature (green) and a skilled forgery (red).}
\label{fig:signals}
\end{figure}

Figure \ref{fig:signals} illustrates one motion signal (accelerometer-$Y$) for a reference signature (black), a genuine signature (green) and a skilled forgery (red).
As can be seen in the figure, the signal values of the reference signature are much closer to those of the genuine signature than to those of the forged signature, and therefore the DTW between the reference signature's signal and the genuine signature's signal is much lower (68.7 vs. 124.93).

\section{Evaluation}
\label{sec:evaluation}

In our evaluation we were interested to answer the following three research questions:
(1) How feasible is our approach to detect forged signatures?
(2) What is the contribution of each motion signal to the task of forgeries detection? and
(3) How robust is our approach to the number of reference signatures used?

\subsection{Experimental Setup}

As common in the field of signature verification, we distinguish between random and skilled forgeries.
In the case of random forgeries, the forger has no information about the user or the user's signature and uses his/her own signature instead.
In this case, the forgery contains a different semantic meaning than the genuine signature provided by the user, presenting a very different overall shape.
In skilled forgeries, the forger has access to the user's name and signature, and often practices forging the user's signature.
This results in forgeries that have stronger resemblance to the genuine signature, and therefore are harder to detect.

Consequently, our evaluation focused on three different verification tasks: discerning between a genuine signature and a skilled forgery, discerning between a genuine signature and a random forgery and finally discerning between a genuine signature and any type of forgery. 

For each user $u$, we randomly select 5 out of her 15 genuine signatures in $G_u$ to serve as her reference set $R_u$ and 8 other signatures as her genuine samples $\overline{G_u}$. 
All of her 15 skilled forgeries in $F_u$ are used as skilled forgeries.
As for random forgeries, we randomly select 10 other users, and sample one random genuine signature of the 15 available.
This entire process of randomized selection of signatures is repeated 25 times with different randomization seeds.

For each of these 25 repetitions, we apply a variation of the known leave-one-out process, where instead of removing one instance at a time, we remove one user at a time.
That is, on each round we use the signatures (genuine and forged) of $65$ users to train our model, and then test the model by using it to classify the signatures (genuine and forged) of the $66$-th user.
This process has two main advantages.
First, our dataset is relatively small, and leave-one-out was shown to work well with small datasets.
Second, our variation simulates a real-world scenario in which we only have a set of reference signatures for the test users (i.e., we do not have additional genuine nor forged signatures for them).

We report the average results over a total of 1650 executions, comprised of 25 repetitions over 66 left-out users.

Four different machine learning models were evaluated to allow for a versatile yet comprehensive representation of model performances.
These models include: na{\"i}ve bayes (a simplified yet fast approach), logistic regression (representing the linear decision boundaries), random forest (representing the more complex boosting models) and neural networks (allowing for extremely complex decision boundaries).
We used the mentioned machine learning algorithms as implemented in Weka \cite{hall2009weka} with their default parameters.

Similar to other studies in this field (e.g., \cite{fischer2015robust}), performance is evaluated in terms of the area under the receiver operating characteristic curve (AUC) and the equal error rate (EER) which is the point in the curve where the false acceptance rate equals the false rejection rate.
(Higher AUC values and lower EER values represent better performing models).

\subsection{Results}

\subsubsection{Verification Accuracy}

\begin{figure}[H]
\centering
\includegraphics[width=0.9\columnwidth]{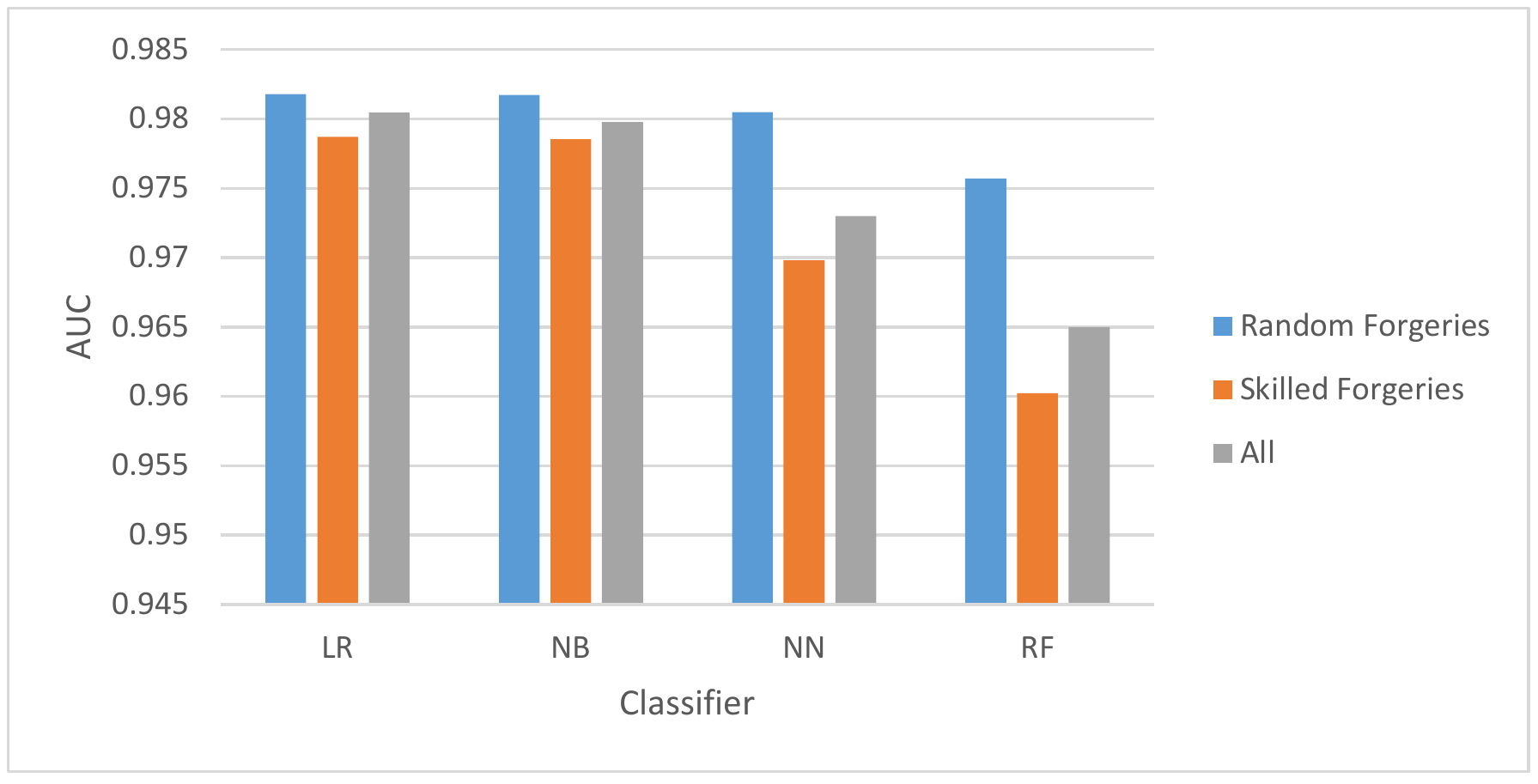}
\caption{Verification performance for different machine learning algorithms and verification tasks - AUC.}
\label{fig:models_auc}
\end{figure}

We first report the results using 5 reference signatures.
The results are summarized in figures \ref{fig:models_auc} and \ref{fig:refs_eer}.

As can be seen from the two figures, the results are quite encouraging for all three verification tasks.
The best results, in all three tasks, were obtained by the Logistic Regression classifier:
AUC=0.978 and EER=0.040 in the skilled forgeries task, 
AUC=0.981 and EER=0.039 in the random forgeries task, 
AUC=0.980 and EER=0.054 in the any forgeries task, 
As expected, our method performed slightly better in the case of random forgeries than in the case of skilled forgeries.

\begin{figure}[H]
\centering
\includegraphics[width=0.9\columnwidth]{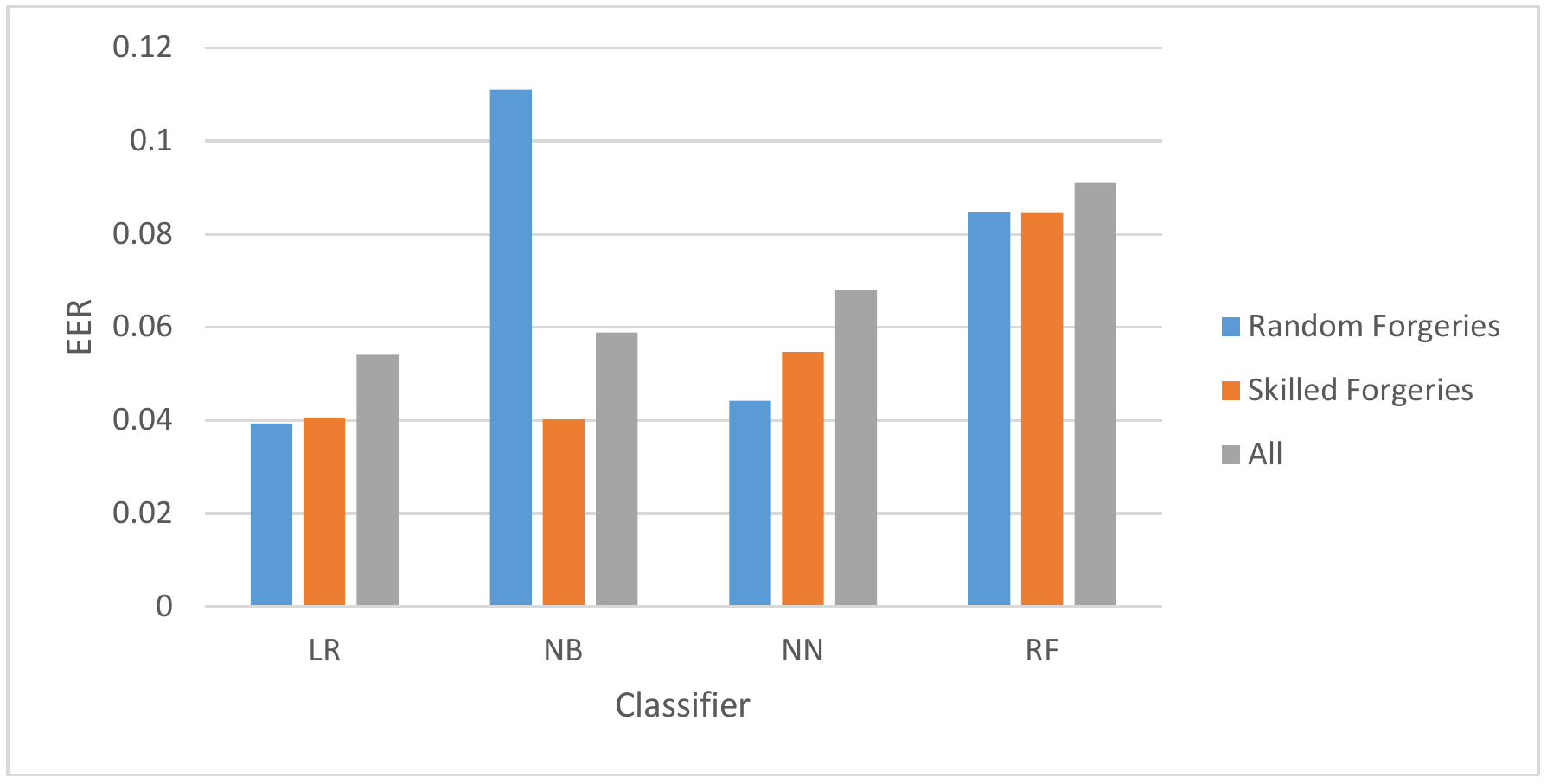}
\caption{Verification performance for different machine learning algorithms and verification tasks - EER.}
\label{fig:models_eer}
\end{figure}

\subsubsection{Features Analysis}

\begin{figure}[H]
\centering
\includegraphics[width=0.9\columnwidth]{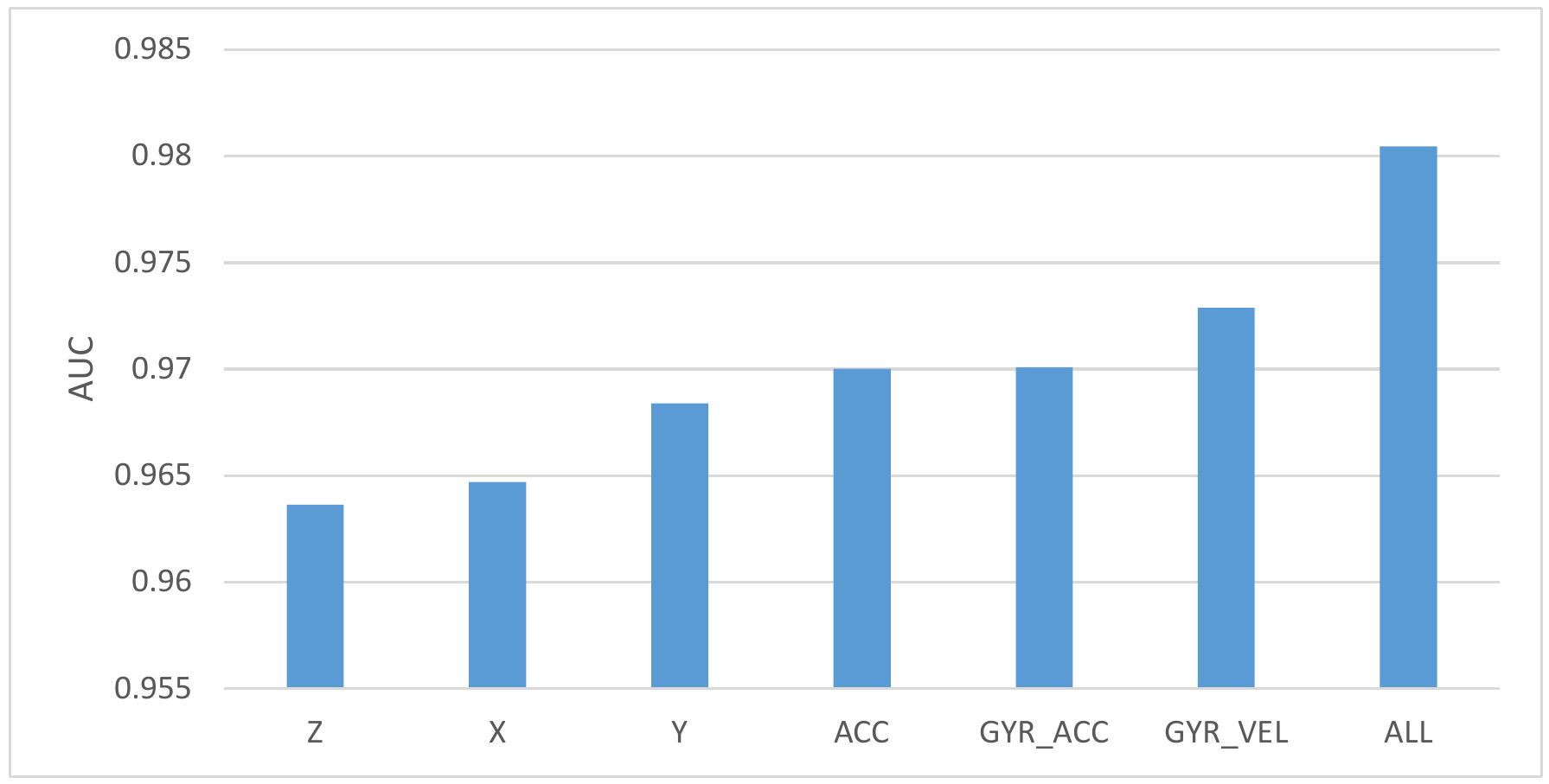}
\caption{Verification performance for different subsets of features - AUC.}
\label{fig:features_auc}
\end{figure}

\begin{figure}[H]
\centering
\includegraphics[width=0.9\columnwidth]{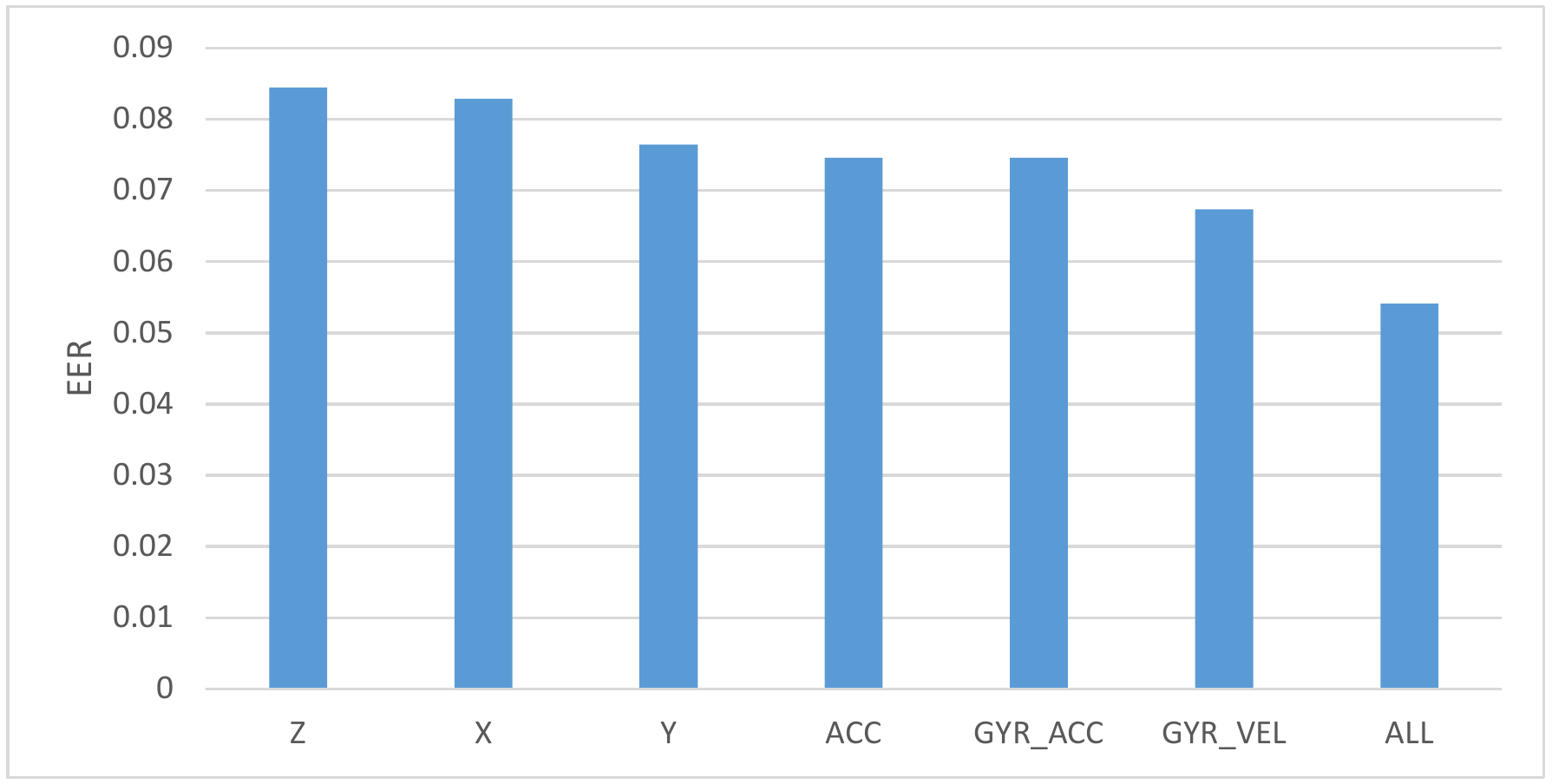}
\caption{Verification performance for different subsets of features - EER.}
\label{fig:features_eer}
\end{figure}

In order to better understand the classification power of the different features we used, we repeated the same evaluation process that was described above, each time using a different subset of features.
More specifically, we used the following seven subsets of features: 1) the three $X$ axis signals, 2) the three $Y$ axis signals, 3) the three $Z$ axis signals, 4) the three Accelerometer signals, 5) the three Gyroscope Acceleration signals, 6) the three Gyroscope Velocity signals and 7) all nine signals.
Figures \ref{fig:features_auc} and \ref{fig:features_eer} report the obtained results (AUC and EER respectively) for these seven subsets of features, focusing on the Logistic Regression classifier and the ``any forgery" verification task.

While we were did not observe major differences between the classification performance of the different subsets of features, minor differences did exist.
For example, we see that the $X$ and $Y$ features preform slightly better than the $Z$ features. 
This makes sense, as the signature process does not require sharp lifting movements of the hand and therefore the $Z$ axis is less significant.
Moreover, we see that the three sensor-based subsets of features (i.e., Accelerometer, Gyroscope Acceleration and Gyroscope Velocity) perform better than the axis-based subsets of features (i.e., $X$, $Y$ and $Z$), implying that all three axes are important for verification.
Finally, combining all nine features together further improves the verification performance.

\subsubsection{How many reference signatures are needed?}

\begin{figure}[H]
\centering
\includegraphics[width=0.9\columnwidth]{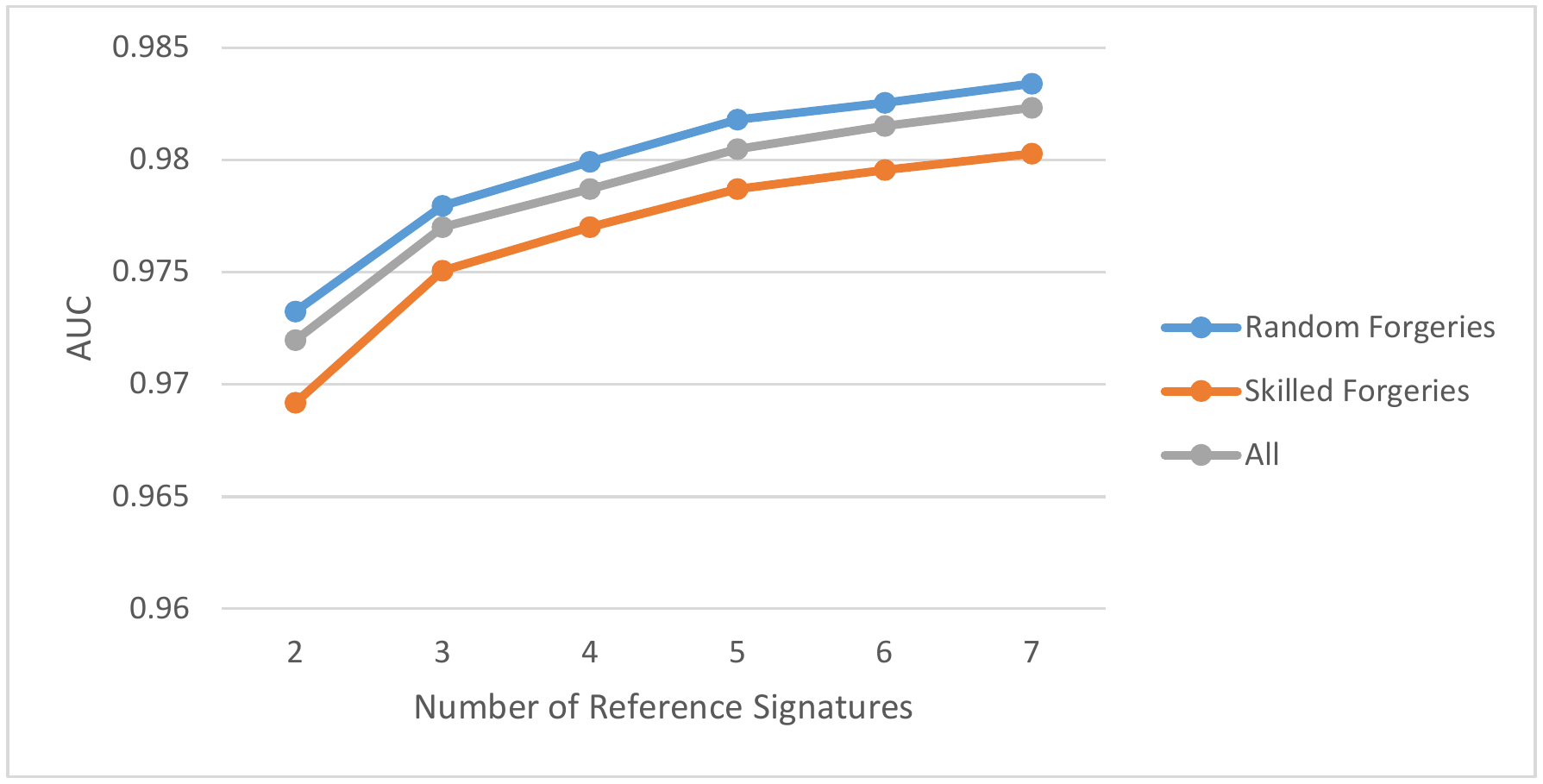}
\caption{Verification performance for different numbers of reference signatures and verification tasks - AUC.}
\label{fig:refs_auc}
\end{figure}

\begin{figure}[H]
\centering
\includegraphics[width=0.9\columnwidth]{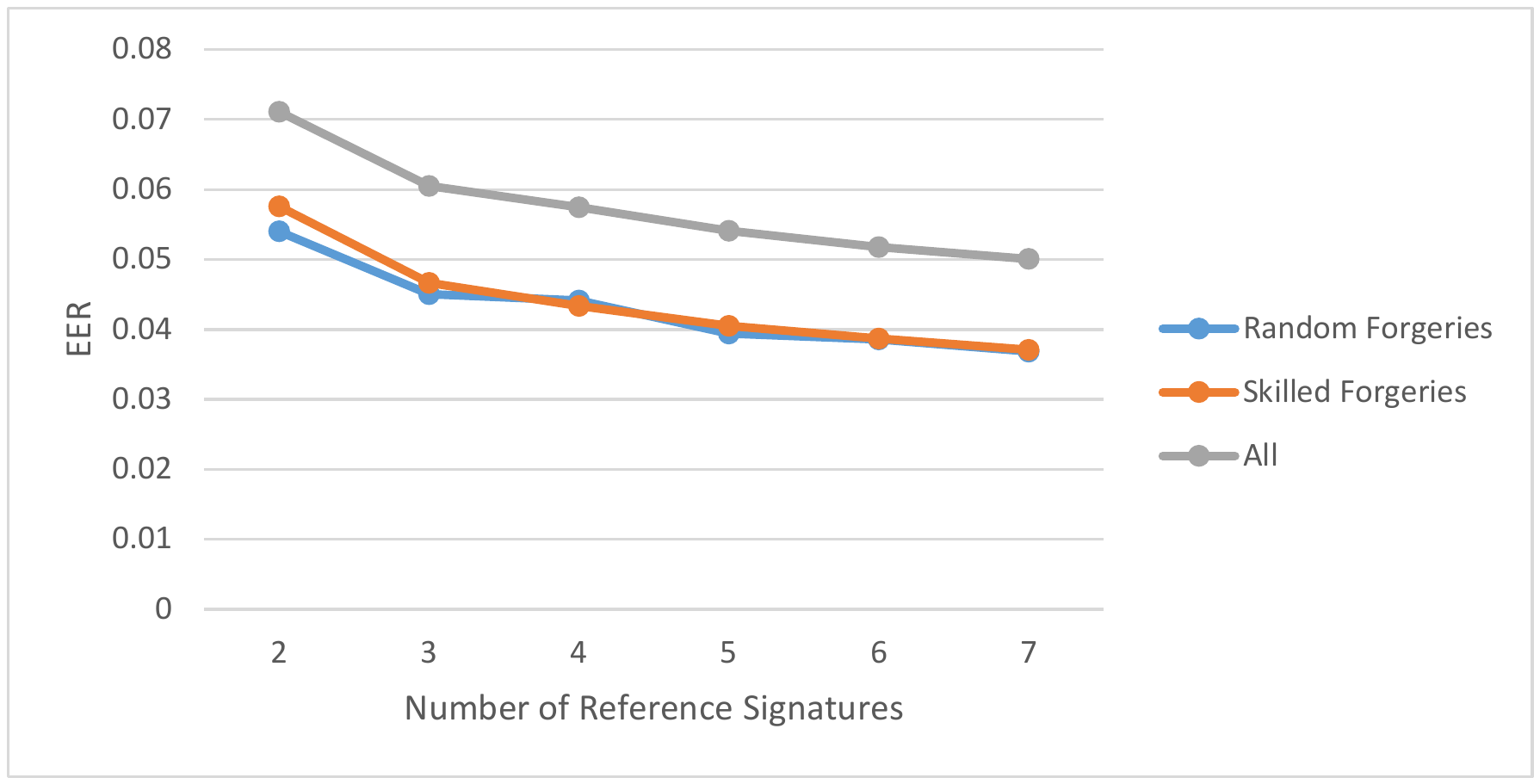}
\caption{Verification performance for different numbers of reference signatures and verification tasks - EER.}
\label{fig:refs_eer}
\end{figure}

In order to test the effect of the size of $R_u$ on the obtained results, we also ran an analysis using varying amounts of reference signatures, ranging between 2 to 7.
Figures \ref{fig:refs_auc} and \ref{fig:refs_eer} report the results for a varying number of reference signatures using the Logistic Regression classifier. 
As can be seen in the figures, our method was able to obtain impressive results, even when only 2 reference signatures were used (AUC=0.969 and EER=0.054).
As expected, the more reference signatures used, the better the results - from AUC=0.969 and EER=0.054 using 2 reference signatures to AUC=0.983 and EER=0.036 using 7 reference signatures.

\section{A Potential Deployment of the System}
\label{sec:deployment}


Imagine a customer, Alice, and a seller, Bob.
Alice is interested in purchasing a product that Bob offers and is planning to pay Bob using a regular check (that she has to fill and sign).
We also assume a trusted third party (e.g., a bank), TTP, that has deployed our verification model.

We suggest the following (skeletal) protocol for signature verification:
\begin{itemize}
\item Step 0: Alice verifies its identity and provides a set of reference signatures to TTP (such as the bank). This step takes place only once for Alice.
\item Step 1: Alice signs-in to a dedicated application of TTP (the application may run on Alice's hand-worn device or on a separate paired device).
\item Step 2: Alice signs the check while wearing the hand-worn device and having the dedicated application running. The application records the movement of the hand during the signing process.
\item Step 3: The application encrypts the signature's recording (using TTP's public-key) and sends the ciphertext to TTP.
\item Step 4: TTP decrypts the ciphertext using its private key.
\item Step 5: TTP retrieves the the set of reference signatures for the signed-in entity.
\item Step 6: TTP executes our verification model on the claimed identity, the recorded signature and the set of reference signatures, and returns the classification result (GENUINE or FORGED) to Bob.
\end{itemize}

Finally, it is worth noting that the protocol described above is not significantly different than existing protocols that banks employ for voice authentication, and therefore suffer from the same type of adversarial attacks (and solutions).
For example, in order to avoid a man-in-the-middle replay attacks, in which an adversary has managed to copy a (ciphertext) recording of the signature, and can use it over and over again, the application can concatenate the signing time to the signature in Step 3, so that TTP will be able to check whether the signature was signed recently in Step 4.
We postpone a more thorough analysis of such potential attacks to the full version of this paper.


\section{Summary and Future Work}
\label{sec:summary}

In this paper, we suggested a new approach for online signature verification that is based on data acquired from hand-worn devices.
In order to evaluate our method, we conducted an experiment involving 66 participants from our institution.
The participants provided both genuine and forged signatures while wearing a hand-worn device equipped with a dedicated software that we developed to record motion data (i.e., accelerometer and gyroscope).
Finally, by analyzing the data, we demonstrated that our method was able to verify signatures with a high degree of accuracy. 

An inherent limitation of our proposed solution is that people must wear the hand-worn device on their dominant hand, which we assume to be the hand they use to sign.
Unfortunately, according to a recent survey (including approx. 4000 subjects) only 34\% of the people wear a watch on their dominant hand \cite{right-left-users-of-watch}.
If we assume this percentage to be valid also in the case of hand-worn devices (especially smartwatches), this implies that 66\% of the people using our system must move the hand-warn device from their non-dominant hand to their dominant hand before signing, therefore making our system less user-friendly.

In future work, we plan to compare our approach with existing state-of-the-art methods for offline and online signature verification.
We would also like to investigate the option of combining data extracted from the wearable device with data collected from a tablet device to achieve even higher verification accuracy.
Such a verification scheme may be found useful in cases where digitizers are already available (e.g., banks), and a wearable device can be added to obtain a higher level of assurance.
Finally, we would like to extract data from additional sensors (besides the accelerometer and gyroscope) such as heart-rate variability and others, and evaluate their impact on verification performance.
The heart-rate sensor, for example, was proven to be useful for detecting lies and is commonly used in lie detector machines.

\bibliographystyle{IEEEtran}
\bibliography{IEEEabrv,sample}

\end{document}